\documentclass{aa}
\usepackage{graphicx}
\def\lae{\mathrel{<\kern-1.0em\lower0.9ex\hbox{$\sim$}}}
\def\gae{\mathrel{>\kern-1.0em\lower0.9ex\hbox{$\sim$}}}
\def\keV{\,{\rm keV} }
\usepackage{epsfig}
\begin{document}
\title{Time resolved spectroscopy of GRB030501 using INTEGRAL\thanks{Based on observations with INTEGRAL, an ESA project
  with instruments and science data centre funded by ESA member states
 (especially the PI countries: Denmark, France, Germany, Italy,
  Switzerland, Spain), Czech Republic and Poland, and with the
  participation of Russia and the USA}}

\author{V. Beckmann\inst{1,}\inst{2}, J. Borkowski\inst{2},
  T.J.-L. Courvoisier\inst{2,3},  D. G\"otz\inst{4}, R. Hudec\inst{5},
  F. Hroch\inst{2,}\inst{5}, N. Lund\inst{6}, S. Mereghetti\inst{4}, S.E. Shaw\inst{2,}\inst{7}, C. Wigger\inst{8}}
\offprints{Volker.Beckmann@obs.unige.ch}
\institute{Institut f\"ur Astronomie und Astrophysik, Universit\"at
 T\"ubingen, Sand 1, D-72076 T\"ubingen, Germany
\and INTEGRAL Science Data Centre, Chemin d' \'Ecogia 16, CH-1290
 Versoix, Switzerland
\and Geneva Observatory, 51 Chemin des Maillettes, CH-1290 Sauverny, Switzerland
\and Istituto di Astrofisica Spaziale e Fisica Cosmica, CNR v. Bassini 15, I-20133 Milano, Italy
\and Astronomical Institute, Academy of Sciences of the Czech
  Republic, CZ-25165 Ondrejov, Czech Republic 
\and Danish Space Research Institute, Juliane Maries Vej 30, DK-2100, Copenhagen, Denmark
\and School of Physics and Astronomy, University of Southampton, Southampton,
SO17 1BJ, United Kingdom
\and Paul Scherrer Institut, CH-5232 Villingen, Switzerland}
\date{Received date; accepted date}
\authorrunning{Beckmann et al.}
\titlerunning{Time resolved spectroscopy of GRB030501 using INTEGRAL}

\abstract{ 
The Gamma-ray instruments on-board INTEGRAL offer an unique opportunity
to perform time resolved analysis on GRBs. The imager IBIS allows accurate positioning of GRBs and broad band spectral analysis, while SPI
provides high resolution spectroscopy. GRB 030501 was discovered by the INTEGRAL Burst
Alert System in the ISGRI field of view. Although the burst was
fairly weak (fluence $F_{20-200 \keV} \simeq 3.5 \times
10^{-6}\,\rm{erg\,cm}^{-2})$ it was possible to perform time resolved
spectroscopy with a resolution of a few seconds.
The GRB shows a spectrum in the 20 - 400 keV range which is consistent
with a spectral index
$\Gamma = -1.7$. No emission line or spectral break was detectable in the spectrum. Although the flux seems to be correlated with the
hardness of the GRB spectrum, there is no clear soft to hard evolution
seen over the duration of the burst. The INTEGRAL data have been
compared with results from the Ulysses and RHESSI experiments.
}
\maketitle

\section{Introduction}
Gamma Ray Bursts (GRBs) were discovered by chance in
the late 1960s by the Vela experiments (Klebesadel et
al. \cite{firstgrb}). They have been proven to be
extragalactic in origin after a successful identification of an X-ray afterglow
by BeppoSAX (GRB970508; Piro et al. \cite{piro}) with an optical counterpart at redshift $z =
0.835$ (Metzger et al. \cite{metzger}). Even though we are now rather
confident that long GRBs are related to massive explosions in distant
galaxies, there are still many open questions remaining. First,
whether GRBs are related to Supernova explosions, and also, what the
connection to the star formation phenomenon is. Another crucial point
is the exact mechanism by which GRBs can produce an energy output of $>
10^{52} \rm ergs$ (under the assumption that the emission is isotropic, which is probably not true). 
Prompt observation of GRBs in several energy ranges is essential to
obtain high quality data for the study of these rapidly fading objects. Although GRBs
were not one of the main targets for the scientific program of INTEGRAL
(Winkler et al. \cite{winkler}), the two main Gamma-ray
instruments, the imager IBIS (Ubertini et al. \cite{ubertini}) and the
spectrometer 
SPI (Vedrenne et al. \cite{vedrenne}), offer great capabilities for studying the
prompt emission of GRBs when they occur in the field of
view. Since the field of view is about 29 degrees, one
Gamma-ray burst per month is expected to be observed. This rate has
been confirmed so far by the six bursts 
in the field of view 
between November 2002 and May 2003:
GRB021125 (Bazzano \& Paizis \cite{grb021125}), GRB021219 (Mereghetti et al. \cite{grb021219}), GRB030131 (Borkowski et al. \cite{grb030131}), GRB030227 (Mereghetti et al. \cite{grb030227}), GRB030320 (Mereghetti et al. \cite{grb030320}), and GRB030501 (Mereghetti et al. \cite{detection}).  In addition, the anticoincidence shield (ACS) of SPI can be used as an all-sky monitor for GRBs (von Kienlin et al. \cite{acs}). 

During the last three bursts in the field of view, both SPI and IBIS, were
in full operational mode, allowing time resolved spectral analysis.

\section{\label{integralobs}INTEGRAL observation}
\begin{figure}
\centering
 \includegraphics[width=6cm,angle=0]{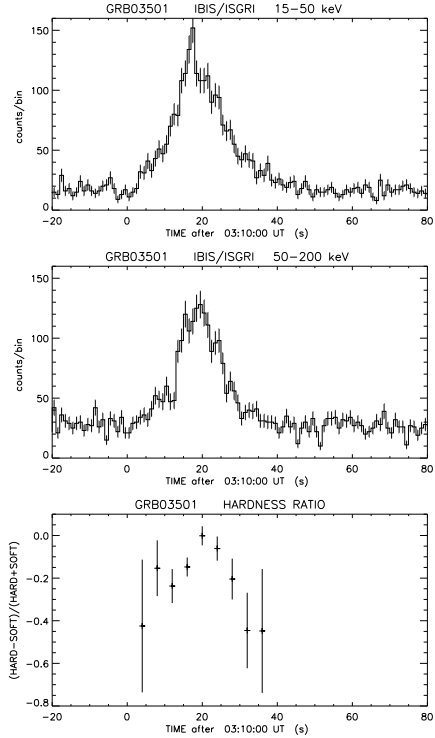}
\caption[]{\label{fig:ibas}Top panels: ISGRI lightcurves in the soft
 and hard band, respectively. IBAS triggered the GRB alert, within half a minute from the burst start, based on the ISGRI data. Bottom panel:
 hardness ratio evolution derived from the ISGRI countrates.}
\end{figure}

GRB030501 was detected on 1$^{\rm{st}}$ May 2003 at 03:10:18 UT with data from the low energy detector of the IBIS instrument, the Integral Soft Gamma Ray Imager
(ISGRI; Lebrun et al. \cite{lebrun}), which consists of $128 \times
128$ CdTe crystals sensitive in the energy range $15 - 300 \keV$. ISGRI uses the coded mask
technique and offers an instrumental resolution of $\sim 12$ arcmin
over the field of view of $29^{\circ} \times
29^{\circ}$. The source location precision depends on the brightness of the source, and
is about 1 arcmin for sources with a detection significance of $10
\sigma$.
The GRB was detected in the ISGRI data by the INTEGRAL Burst Alert System (IBAS;
Mereghetti et al. \cite{ibas}), which automatically determines the
position and time of events which occur in the IBIS field of view. The IBAS
alert was distributed approximately 30 seconds after the burst start
time with a positional uncertainty of only 4 arcmin (Mereghetti et al. \cite{detection}). The ISGRI lightcurve, is shown in Fig. \ref{fig:ibas}.

A later analysis, performed off-line, on the ISGRI data revealed a
position of (J2000.0) $\rm 19^h 05^m 30^s$, $\rm +6^\circ 18^m 26^s$
with an uncertainty of $\sim3$ arcmin
(Mereghetti et al. \cite{detection}). The Gamma-Ray burst had a duration of 40 seconds.
Data from the spectrometer SPI were also analysed at
the INTEGRAL Science Data Centre (ISDC; Courvoisier et al. \cite{thierry}, Beckmann \cite{beckmann}). SPI
is designed for high spectral resolution (FWHM of 2.5 keV at 1 MeV) in
the energy range $20 - 8000 \keV$, which is achieved by 19 cooled Germanium detectors. A
coded mask with 128 elements provides an instrumental spatial
resolution of 2.8 degrees. The position accuracy for point sources can
be $< 5$ arcmin (for a source with $S/N = 100$).
The position for GRB030501 extracted from the SPI data is (J2000.0)
$\rm 19^h 06^m$, $\rm +7^\circ 6^m$ ($\pm 1^\circ$), which is about 50
arcmin off the position detected by ISGRI. As the GRB falls into the
partially coded field of view, the low position accuracy of SPI is not
surprising, as only part of the detector plane (5 out of 19 detectors)
is in fact illuminated by the event. 

\begin{figure}
\centering
 \includegraphics[width=6.0cm,angle=-90]{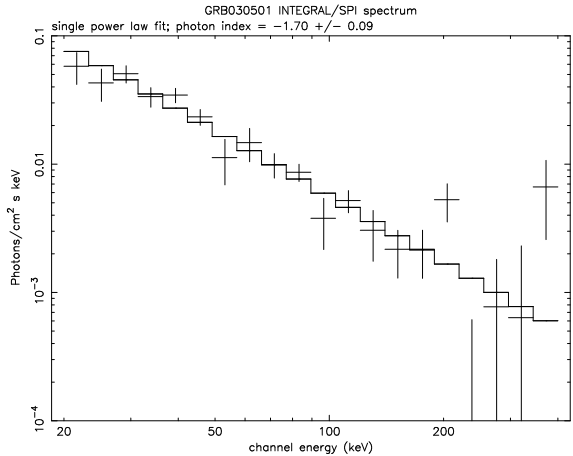}
\caption[]{\label{fig:spectrum}GRB spectrum
  in the range 20 - 400 keV, taken from the SPI data over 20 seconds after the burst occurance.} 
\end{figure}

The peak gamma ray flux, $f_{20-200 \keV}$, in the 20 - 200 keV band
measured by SPI in a 2 second bin starting at 03:10:20 UTC is $\rm 2.8
\pm 0.4 \,photons\, cm^{-2}\, sec^{-1}$.  This is consistent with
ISGRI where $f_{20-200 \keV} = \rm
2.7 \pm 1.2 \,photons\, cm^{-2}\, sec^{-1}$ was measured in a 1 second
bin starting at 03:10:18.4 UTC, reaching the peak about 1 second
before it occurs in the SPI data. The fluence, $F_{20-200 \keV}$,
measured for the GRB by both instruments in the same band and
integrated over the full burst visibility period is also consistent:  
\begin{eqnarray*}
{\rm SPI}:   F_{20-200 \keV} & = & 39.3 \pm 2.5 \rm{\,\,photons\,cm}^{-2}\\   
                  & = & 3.7 \pm 0.2 \times 10^{-6}\,\rm{erg\,cm}^{-2} \\
{\rm ISGRI}: F_{20-200 \keV} & = & 37.5\pm 12.5 \rm{\,\,photons\,cm}^{-2}\\ 
                  & = & 3 \pm 1 \times 10^{-6}\,\rm{erg\,cm}^{-2}
\end{eqnarray*}
The GRB was also observed by the Ulysses experiment (Hurley et al. \cite{hurley}).  Due to the weakness of the detection the reported fluence, in the 25 - 100 keV band, is uncertain by about a factor of two, but this is still consistent with the measurements made by the INTEGRAL instruments:  
\begin{eqnarray*}
{\rm SPI}:   F_{25-100 \keV}    & \simeq & 2.0 \times 10^{-6} \, \rm{erg \, cm}^{-2}\\
{\rm ISGRI}:  F_{25-100 \keV}   & \simeq & 1.5 \times 10^{-6} \, \rm{erg \, cm}^{-2}\\
{\rm ULYSSES}:  F_{25-100 \keV} & \simeq & 1.1 \times 10^{-6}\,\rm{erg\,cm}^{-2}
\end{eqnarray*}
The $\sim 30 \%$ uncertainty on ISGRI fluence is dominated by systematic errors on
the response of the instrument at large off-axis angles. We note however
a good agreement with the SPI value, which confirms the value obtained
with ISGRI.
The overall spectrum of the burst based on SPI data is shown in Figure
\ref{fig:spectrum}. The GRB occurred in a pointed observation of 1800 sec
length. The background emission was estimated from this pointing, but
excluding the time when the GRB occurred. The GRB is detectable up to
at least 200 keV in both the ISGRI and SPI\footnote{A detailed
description how to perform the extraction of a GRB from SPI data can
be found at
http://isdc.unige.ch/Instrument/spi/pages/usermanual.html} data. A single
power law represents the SPI data well, resulting in a photon index of
$\Gamma = -1.70 \pm 0.09$.  A more complicated model (e.g. a broken power
law or a Band model; Band et al. \cite{band}) does not improve the fit, thus no spectral break is
 detectable. This result is consistent with the spectral slope of
 $\Gamma = -1.75 \pm 0.10$ derived from ISGRI data. For the ISGRI data
 the background can be estimated at the same time as the source flux,
 using the Pixel Illumination Function (PIF). Spectra are extracted
 computing one PIF for each energy bin (128 linearly spaced bins have
 been used).\\

There was a marginal detection of the GRB by the SPI-ACS (Hurley et
al. \cite{hurley}). The low countrate of this GRB is expected in the ACS data, as the effective area for events in the field of view of SPI is small for the ACS, which shields the spectrograph from the
sides and from its back (von Kienlin et al. \cite{acs}). The combination of Ulysses and
INTEGRAL data also allowed triangulation of the GRB event by the 3$^{\rm{rd}}$
Interplanetary Network (IPN). The result is consistent with the ISGRI
position (Hurley et al. \cite{hurley}).
The INTEGRAL X-ray (JEM-X) and optical (OMC) monitors were unable to provide any additional information since the GRB was located well outside of the respective fields of view of these instruments.

\section{Comparison with RHESSI observation of GRB030501}
GRB 030501 has also been seen by the spectrometer of the Ramaty High
Energy Solar Spectroscopic Imager (RHESSI), which is a NASA Small Explorer satellite designed to study hard X-rays 
and gamma-rays from solar flares (Lin et al. \cite{HESSI}).
The instrument consists of 9 germanium detectors,
each of volume $300 \, \rm cm^{3}$, that cover an
energy range of 3 keV to 17 MeV, with an energy resolution
of about 3 keV (FWHM) at 1 MeV (Smith et al. \cite{HESSI2}). 
The detector uses a Rotation Modulation Collimator (RMC) system 
for high resolution imaging of solar flares. 
The germanium detectors are only lightly shielded.
Above about 60 keV, they have a significant response to photons
from any direction in the sky. 
Thus, RHESSI is a sensitive GRB detector,
and as such it is part of the IPN.

\begin{figure}
\begin{center}
\epsfxsize=7.0cm
\epsfbox{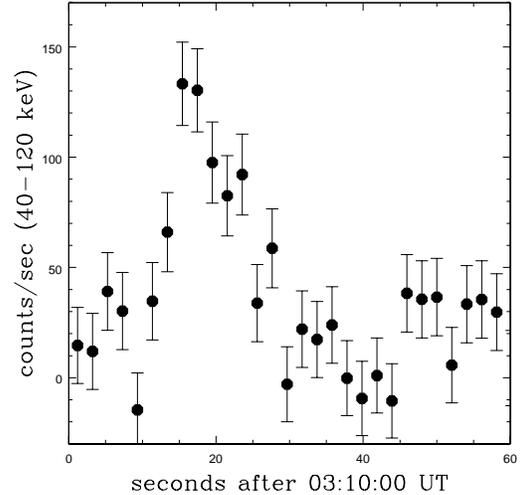}
\caption[]{\label{fig:lightcurvehessi}Lightcurve of GRB 030501 measured by the spectrograph of RHESSI.} 
\end{center}
\end{figure}

The lightcurve of GRB030501 as seen by RHESSI in the $40 - 120$ keV band
is shown in Fig. \ref{fig:lightcurvehessi}. The peak photon flux measured is
$f_{70-200 \rm keV} \simeq
0.55 \pm 0.17 \rm \,photons\, cm^{-2} \, sec^{-1}$, and the fluence 
over the 20 sec burst duration 
is $F_{70-200 \rm keV} \simeq 2.1 \pm 0.6 \times 10^{-6} \,
\rm erg \, cm^{-2}$.   


\section{Time resolved results on GRB030501}
\begin{figure}
\begin{center}
\epsfxsize=7.0cm
\epsfbox{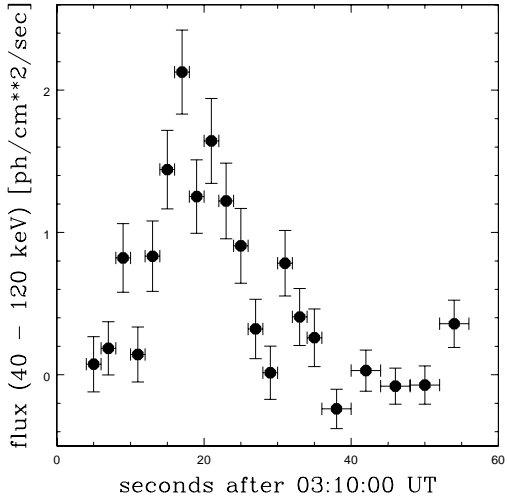}
\caption[]{\label{fig:lightcurve}Lightcurve of GRB 030501 in the same
  energy band as the RHESSI data, taken from INTEGRAL/SPI} 
\end{center}
\end{figure}

Although the burst is comparably weak, the sensitivity of ISGRI and SPI allows the
study of the lightcurve of the prompt emission. We show the SPI 
lightcurve in the same energy band as for RHESSI in Fig. \ref{fig:lightcurve}. The peak flux is reached $\sim 10$ seconds
after the burst started.
Spectra were extracted from the SPI data, in 5 logarithmically binned
energy bands between 20 and 400 keV. For ISGRI the data have been
binned in order to have at least 25 counts per bin. XSPEC 11.2 was used to fit a single power law to the data in time bins of $2 - 10$
seconds over a period of 30 seconds after the burst started.

\begin{figure}
\begin{center}
\epsfxsize=7.0cm
\epsfbox{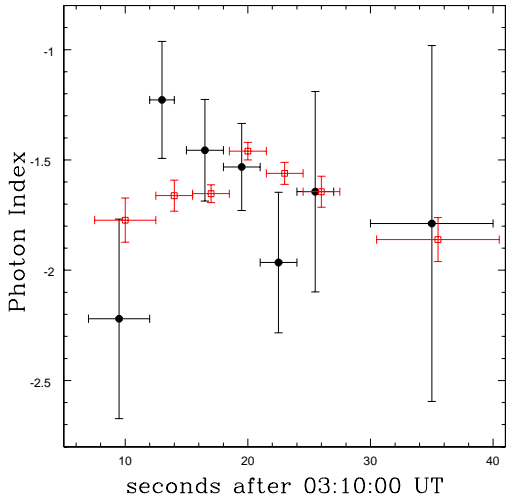}
\caption[]{\label{fig:evolution}Evolution of the spectral slope of a single power law,
  fitted to the ISGRI (open squares) and SPI (filled circles) data of
  GRB 030501. The ISGRI data points have been shifted by $+0.5$ sec
  for better visibility. The definition of the
  photon index is $f_{\nu} \propto \nu^{\Gamma}$.}
\end{center}
\end{figure}

The results are shown in Fig. \ref{fig:evolution}. 
The spectrum starts apparently rather steep, but
then immediately has a photon index of $\Gamma \simeq -1.5$ as the flux
increases. Before the GRB is
below the instrumental sensitity, it apparently softens again. 
In the ISGRI data there is evidence of hardness intensity
correlation as seen in other GRBs before (e.g. Ford et
al. \cite{ford}). This is consistent with the SPI data,
though the statistic is not high enough to constrain the results. 
No clear spectral evolution is seen in the data. 
The hardness ratio evolution in the RHESSI data show a similar trend
to the one seen in ISGRI (Fig. \ref{fig:ibas}), though the error bars are
larger.

\section{Discussion}
Optical follow up observation has not revealed an optical counterpart
to this GRB (Boer \& Klotz \cite{boer}; Fox \cite{fox}; Rumjantsev et
al. \cite{rumyantsev}). Using the
1m Wise telescope in comparison with POSS-II E photographic plates,
the magnitude of the optical counterpart 16.5 hours after the prompt
emission can be limited to $R \ge 20.0 \rm \, mag$ (Ofek et
al. \cite{ofek}).
Also analysis of optical observations carried out with the automatic $25
\rm \, cm$ TAROT telescope shortly after the burst
occurrence (i.e. within 15 minutes) shows no optical counter part with
$R\le 18.0 \rm \, mag$ (Klotz \& Boer \cite{klotz}).

With a Galactic latitude
of only 0.2 degree and an estimated extinction of $E(B-V) \sim 15$, identification of an optical counterpart is indeed difficult, if not impossible.


As no break in the spectrum was detected in either the SPI
or the ISGRI data, it can be assumed that the peak energy of this
long GRB is either $E_{peak} \lae 30 \keV$ or $E_{peak} \gae 200
\keV$. Since a very low peak energy is rather unlikely (see Preece et
al. \cite{preece}), we assume a spectral break above 200 keV. 
GRB030501 shows a similar spectral behaviour to bursts studied before
(e.g. GRB921207; Ghirlanda et al. \cite{ghirlanda}) but is
about a factor $\sim 10$ weaker than the bursts where time resolved
spectroscopy has been possible with data from previous missions.


The comparison with the RHESSI data shows that this experiment
is also a powerful
tool in the detection and spectral analysis of GRBs. 
Especially for GRBs, which are not in the field of view of the INTEGRAL
main instruments SPI and IBIS, RHESSI provides sufficient spectral
and timing resolution ($16 \mu$sec) to study those events, as the
RHESSI spectrograph is a non-shielded all-sky monitoring instrument.

This GRB demonstrates the great capabilities of INTEGRAL and the
software package, provided by the ISDC in collaboration with the
instrument teams. The time lag between GRB occurrence and providing
detailed spectral and timing analysis is less than half a day. 


\begin{acknowledgements}
RH and FH acknowledge the support provided by the ESA Prodex Project
14527. JB was supported by the Polish grant 2P03C00619p02 from KBN and
SS by PPARC grant GR/K/94867.
\end{acknowledgements}

\end{document}